\documentclass[conference,letterpaper,final]{IEEEtran}

\usepackage{hyperref}
\usepackage{cite}

\usepackage[dvips]{graphicx}
\graphicspath{{../figures/}}
\DeclareGraphicsExtensions{.eps}

\usepackage[cmex10]{amsmath}
\usepackage{amssymb,amsthm}
\usepackage{dsfont}
\usepackage{setspace}

\usepackage{mathtools}
\usepackage[caption=false,font=footnotesize,labelfont=sf,textfont=sf]{subfig}
\newtheorem{remark}{Remark}
\newtheorem{lemma}{Lemma}
\newtheorem{theorem}{Theorem}
\newtheorem{corollary}{Corollary}

\newtheorem{proposition}{Proposition}

\usepackage{psfrag}
\usepackage{array}
\usepackage{mdwmath}
\usepackage{mdwtab}
\usepackage{cases}
\usepackage{enumerate}
\usepackage{eqparbox}

\usepackage{mathtools}

\usepackage{stfloats}

\newcommand{\sir}{\mathtt{SIR}}
\newcommand{\sig}{\mathsf{P}}
\newcommand{\interc}{\mathsf{J}_{\mathcal{C}_{\bar{\text{a}}}}}
\newcommand{\internc}{\mathsf{J}_{\mathcal{\bar C}}}
\newcommand{\intert}{\mathsf{\tilde{J}}}
\newcommand{\lap}{\mathcal{L}}
\newcommand{\pc}{\mathtt{P}_{\text{c}}}
\renewcommand{\th}{^\text{th}}
\newcommand{\lnu}{\lfloor\nu\rfloor}
\newcommand{\unu}{\lceil\nu\rceil}

\IEEEoverridecommandlockouts

\begin{document}
%
\title{Analysis of Non-Coherent Joint-Transmission Cooperation in Heterogeneous Cellular Networks}

\author{Ralph Tanbourgi\IEEEauthorrefmark{1}, Sarabjot Singh\IEEEauthorrefmark{2}, Jeffrey G. Andrews\IEEEauthorrefmark{2} and
Friedrich K. Jondral\IEEEauthorrefmark{1}\thanks{\IEEEauthorrefmark{1}The authors are with the Communications Engineering Lab, Karlsruhe Institute of Technology, Germany. Email: \texttt{ralph.tanbourgi@kit.edu, friedrich.jondral@kit.edu}. This work was supported by the German Research Foundation (DFG) within Priority Program 1397 "COIN" under grant No. JO258/21-1.\newline\IEEEauthorrefmark{2}The authors are with the Wireless and Networking Communications Group (WNCG), The University of Texas at Austin, TX, USA. Email: \texttt{sarabjot@utexas.edu, jandrews@ece.utexas.edu}}
}

\maketitle

\begin{abstract}
Base station (BS) cooperation is set to play a key role in managing interference in dense heterogeneous cellular networks (HCNs). Non-coherent joint transmission (JT) is particularly appealing due to its low complexity, smaller overhead, and ability for load balancing. However, a general analysis of this technique is difficult mostly due to the lack of tractable models. This paper addresses this gap and presents a tractable model for analyzing non-coherent JT in HCNs, while incorporating key system parameters such as user-centric BS clustering and channel-dependent cooperation activation. Assuming all BSs of each tier follow a stationary Poisson point process, the coverage probability for non-coherent JT is derived. Using the developed model, it is shown that for small cooperative clusters of small-cell BSs, non-coherent JT by small cells provides spectral efficiency gains without significantly increasing cell load. Further, when cooperation is aggressively triggered intra-cluster frequency reuse within small cells is favorable over intra-cluster coordinated scheduling.
\end{abstract}


\IEEEpeerreviewmaketitle

\section{Introduction}
The rapid increase of mobile traffic---primarily driven by data-intense applications such as video streaming and mobile web\cite{cisco13}---requires new wireless architectures and techniques. HCNs have attracted much interest due to their potential of improving system capacity and coverage with increasing density. Because of the opportunistic and dense deployment with sometimes limited site-planning, HCNs have at the same time contributed to rendering interference the performance-limiting factor \cite{GhoAnd12}. Base station (BS) cooperation, which aims at increasing the signal-to-interference ratio ($\sir$) at victim users, is a promising technique to cope with newly emerging interference situations.

\subsection{Related Work and Motivation}
BS cooperation has been thoroughly analyzed in \cite{gesbert10,irmer11,lozano12,lee12,barbieri12,marsch11}. To address interference issues associated with heterogeneous deployments and to make use of the increased availability of wireless infrastructure, BS cooperation was also studied for HCNs. In \cite{lee12} the authors demonstrated that with low-power BSs irregularly deployed inside macro-cell coverage areas, BS cooperation achieves higher throughput gains compared to the macro-cell only setting, and hence as HCNs create new and complex cell borders more users profit from tackling other-cell interference through BS cooperation. The applicability of coordinated scheduling/beamforming (CS/CB) cooperation for HCNs was studied in \cite{barbieri12}, where it was found that practical issues such as accurate CSI feedback and tight BS synchronization required for coherent cooperation may disenchantingly limit the achievable gains. Such practical challenges associated with BS cooperation are by no means unique to HCNs \cite{marsch11}, and hence other techniques with less stringent requirements have been studied as well. One such technique is non-coherent JT, in which a user's signal is transmitted by multiple cooperating BSs without prior phase-mismatch correction and tight synchronization across BSs. At the user, the received signals are non-coherently combined, thereby providing opportunistic power gains. The standardization interest for non-coherent JT \cite{3gpp_tr_36819,ericsson11}, is particularly due to its \emph{lower implementation complexity} for both the backhaul and the CSI feedback \cite{li12} and its ability for \emph{balancing load} \cite{barbieri12}; features of essential importance in HCNs \cite{andrews13}.

Besides, analyzing BS cooperation in HCNs entails several challenges due to the many interacting complex system parameters, e.g., radio channel, network geometry, and interference. To make things even more difficult, these parameters typically differ across tiers, e.g., BS transmit power, channel fading or cell association. 
To address these challenges, \emph{stochastic geometry} \cite{stoyan95,ganti09,HaenggiBook} has recently been proposed and used for analyzing cooperation in cellular networks \cite{huang11,keeler13,baccelli13_1,tanbourgi13_3,haenggi13_comp}.

\subsection{Contributions}
In this paper, we model and analyze non-coherent JT cooperation in HCNs. The contributions are summarized below.

\textbf{Analytical model:} A tractable model for HCNs with non-coherent JT is proposed in Section~\ref{sec:model}. The model incorporates cooperation aspects of practical importance such as user-centric clustering and channel-dependent cooperation activation, each of which with a tier-specific threshold that models the complexity and overhead allowed in each tier. Other aspects such as BS transmit power, path loss, and arbitrary fading distribution are also assumed tier-specific.

\textbf{Coverage probability:} As the main result, the coverage probability under non-coherent JT is characterized in Section~\ref{sec:cov_prob} for a typical user. The main result has a compact semi-closed form (derivatives of elementary functions) and applies to general fading distributions. We also propose a simple but accurate linear approximation of the coverage probability.   

\textbf{Design insights:} 
\textit{Load balancing:} Balancing load in two-tier HCNs, by additionally pushing more users to small cells in order to let these cells assist macro BSs with non-coherent JT, is favorable only to a limited extent. As small-cell cooperative clusters are increased, spectral efficiency gains grow only approximately logarithmically while cell load in those cells increases much faster. At small cluster sizes of small cells, generously stimulating cooperation by channel-dependent cooperation activation yields considerable spectral efficiency gains without consuming much radio resources.\\
\textit{Intra-cluster scheduling in small cells:} When cooperation is aggressively triggered, small cells should reuse the resources utilized by non-coherent JT, i.e., intra-cluster frequency reuse (FR), to obtain cell-splitting gains. In lightly-loaded small cells with less aggressive triggering, not reusing these resources, i.e., intra-cluster CS,  is better to avoid harmful interference.

\section{Mathematical Model}\label{sec:model}

\subsection{Heterogeneous Network Model}
\begin{figure}[t]
	\psfrag{tag1}[c][c]{\small{$\sim\hspace{-.05cm}\Delta_{1}$}}
	\psfrag{tag2}[c][c]{\small{$\sim\hspace{-.07cm}\Delta_{2}$}}
	\psfrag{tag3}[c][c]{\small{Cooperative Tier-2 BSs}}
	\psfrag{tag4}[c][c]{\small{Cooperative Tier-1 BSs}}
	   \centering
    \includegraphics[width=0.47\textwidth]{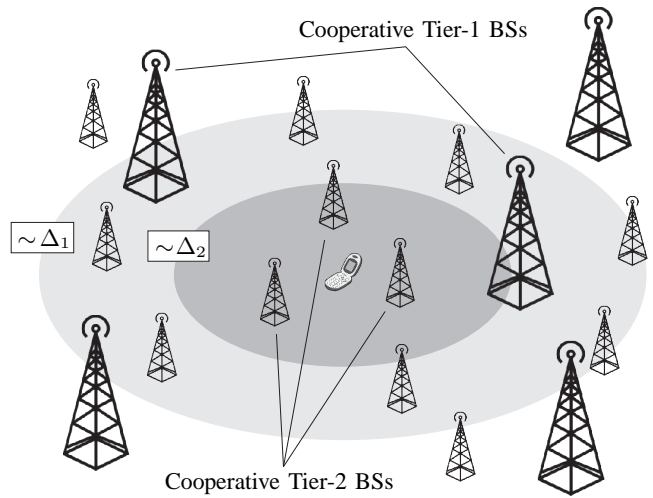}
\caption{Illustration of the considered scenario for the example of a two-tier cooperative HCN: Tier-1 BSs in range (inside lightly-shaded region with radius~$\sim\Delta_1$) form a cooperative cluster for the typical user. Nearby Tier-2 BSs (inside dark-shaded region with radius~$\sim\Delta_2$) join this cooperative cluster. All other nodes create out-of-cluster interference.}\vspace{-.2cm}
\label{fig:illustration}
\end{figure}

We consider an OFDM-based co-channel $K$-tier HCN with single-antenna BSs in the downlink. The locations of the BSs in the $k\th$ tier are modeled by a stationary planar Poisson point process (PPP) $\Phi_{k}$ with density $\lambda_{k}$. The BS point processes $\Phi_{1},\ldots,\Phi_{K}$ are assumed independent. Every BS belonging to the $k\th$ tier transmits with power $\rho_{k}$. A signal transmitted by a $k\th$ tier BS undergoes a distance-dependent path loss $\|\cdot\|^{-\alpha_{k}}$, where $\alpha_{k}>2$ is the path loss exponent of the $k\th$ tier. Fig.~\ref{fig:illustration} illustrates the considered scenario.

The entire set of BSs, denoted by $\Phi$, is formed by superposition of the individual random sets $\Phi_{k}$, i.e., $\Phi\triangleq\bigcup_{k=1}^{K}\Phi_{k}$. By \cite{stoyan95}, the point process $\Phi$ is again a stationary PPP with density $\lambda=\sum_{k}\lambda_{k}$. We assume single-antenna users/receivers to be distributed according to a PPP. By Slivnyak's Theorem\cite{stoyan95}, we evaluate the system performance at a \emph{typical} receiver located at the origin without loss of generality.

The transmitted signals are subject to (frequency-flat) block-fading. The (power) fading gain from the $i$-th BS in the $k\th$ tier to the typical user at the origin is denoted by $\mathsf{g}_{ik}$. We assume that the $\mathsf{g}_{1k},\mathsf{g}_{2k},\ldots$ are i.i.d., i.e., the fading statistics may possibly differ across the $K$ tiers. When appropriate, we will drop the index $i$ in $\mathsf{g}_{ik}$. We further require that $\mathbb{E}\left[\mathsf{g}_{ik}\right]=1$ and $\mathbb{E}\left[\mathsf{g}_{ik}^2\right]<\infty$ for all $i,k$. Heterogeneous propagation conditions might, for instance, be due to  different antenna heights across tiers. Thermal noise is neglected for analytical tractability but can be included in the analysis \cite{tanbourgi13_3}.

\subsection{Non-Coherent Cooperation Model}

\textit{BS clustering model:} We employ a dynamic user-centric BS clustering method. 
In this method, BSs with sufficiently high average received signal strength (RSS) monitored at a given user form a cooperative cluster to cooperatively serve this user.\footnote{Practical constraints typically impose additional criteria to this simple clustering rule for the associated overhead not to be overwhelming, cf. \cite{marsch11} for an elaborate discussion. We leave such possible extensions for future work.} Transferring this to the model, the $i$-th BS from the $k\th$ tier at location $\mathsf{x}_{ik}$ belongs to the cooperative cluster of the typical user if $\rho_{k}\|\mathsf{x}_{ik}\|^{-\alpha_{k}}\geq \Delta_{k}$. Hereby, $\Delta_{k}$ denotes the $k\th$ tier RSS threshold, which depends on the allowable cooperation overhead in the $k\th$ tier and serves as a design parameter. The set of cooperative BSs from the $k\th$ tier, then, has the form
\begin{IEEEeqnarray}{rCl}
	\mathcal{C}_{k}\triangleq\left\{\mathsf{x}_{ik}\in\Phi_{k}\left|\right.\|\mathsf{x}_{ik}\|\leq\left(\tfrac{\Delta_{k}}{\rho_k}\right)^{-1/\alpha_k}\right\}\label{eq:cluster}.
\end{IEEEeqnarray}
The corresponding subset of non-cooperative BSs is denoted by $\mathcal{\bar C}_{k}\triangleq \Phi_{k}\setminus\mathcal{C}_{k}$.


\begin{remark}
Practical user-centric clustering methods slightly differ from the above clustering model as the RSS \textit{difference} to the serving BS is considered. Modeling this kind of clustering is analytically more involved and is deferred to future work.
\end{remark}

\newcounter{mycounter}
\begin{figure*}[!t]
\normalsize
\setcounter{mycounter}{\value{equation}}
\setcounter{equation}{7}
\begin{IEEEeqnarray}{rCl}
	\mathcal{L}_{\sig_k}(s)&=&\exp\hspace{-.05cm}\left\{\hspace{-.05cm}-\lambda_{k}\pi\rho_{k}^{2/\alpha_k}\mathbb{E}_{\mathsf{g}_k}\hspace{-.12cm}\left[\max\hspace{-.05cm}\left\{\Delta_k,\tfrac{T_k}{\mathsf{g}_k}\right\}^{\hspace{-.07cm}-2/\alpha_k}\hspace{-.14cm}\left(1-e^{-s\mathsf{g}_{k}\max\{\Delta_k,T_k/\mathsf{g}_k\}}\right)\hspace{-.07cm}+\hspace{-.07cm}(s\mathsf{g}_k)^{2/\alpha_k}\Gamma(1-\tfrac{2}{\alpha_k},s\mathsf{g}_k\hspace{-.05cm}\max\{\Delta_k,\tfrac{T_k}{\mathsf{g}_k}\}\right]\right\}\IEEEeqnarraynumspace\label{eq:lap_pk}
\end{IEEEeqnarray}
\setcounter{equation}{\value{mycounter}}
\hrulefill
\vspace*{4pt}
\end{figure*}

\textit{Channel-dependent cooperation activation:} Whether a BS of a cooperative cluster gets engaged in a cooperative transmission to a particular user typically depends on its instantaneous channel to that user. To capture the basic impact of this channel-dependent mechanism, we use the following model: the $i$-th cooperative BS of the $k\th$ tier joins a cooperative transmission to the typical user if $\mathsf{g}_{ik}\rho_{k}\|\mathsf{x}_{ik}\|^{-\alpha_{k}}\geq T_{k}$, where $\mathsf{x}_{ik}\in\mathcal{C}_{k}$ and $T_{k}$ is the cooperation activation threshold corresponding to the $k\th$ tier. Similar to $\Delta_{k}$, the variable $T_{k}$ serves as a tunable design parameter to trade off performance against overhead. The subset of \emph{active} cooperative BSs from the $k\th$ tier serving the typical user is denoted as
\begin{IEEEeqnarray}{rCl}
	\mathcal{C}_{\text{a},k}\triangleq\left\{\mathsf{x}_{ik}\in\mathcal{C}_{k}\left|\right.\|\mathsf{x}_{ik}\|\leq \left(\tfrac{T_{k}}{\mathsf{g}_{ik}\rho_k}\right)^{-1/\alpha_k}\right\}.\label{eq:scheduled}
\end{IEEEeqnarray}
We denote by $\mathcal{C}_{\bar{\text{a}},k}\triangleq \mathcal{C}_{k}\setminus\mathcal{C}_{\text{a},k}$ the set of cooperative BSs from the $k\th$ tier not participating in the cooperative transmission to the typical user. These BSs may remain silent (intra-cluster CS) or may serve other users (intra-cluster FR) on the resources used for the cooperative transmission.

\textit{Non-coherent joint-transmission:} In non-coherent JT, BSs scheduled for cooperative transmission to a user transmit the same signal without prior phase-alignment and tight synchronization to that user. At the user, the multiple copies are received non-coherently. At the typical user, the $\sir$ can then be expressed as \cite{tanbourgi13_3}
\begin{IEEEeqnarray}{rCl}
	\sir &\triangleq& \frac{\sig}{\interc+\internc},\label{eq:sir}\IEEEeqnarraynumspace
\end{IEEEeqnarray}
where
\begin{itemize}
	\item $\sig\triangleq\sum_{k}\sum_{\mathsf{x}_{ik}\in\mathcal{C}_{\text{a},k}}\mathsf{g}_{ik}\rho_{k}\|\mathsf{x}_{ik}\|^{-\alpha_{k}}$ is the received signal power,
	\item $\interc\triangleq \sum_{k}\sum_{\mathsf{x}_{ik}\in\mathcal{C}_{\bar{\text{a}},k}}\mathsf{g}_{ik}\rho_{k}\|\mathsf{x}_{ik}\|^{-\alpha_k}$ is the intra-cluster interference,
	\item $\internc\triangleq\sum_{k}\sum_{\mathsf{x}_{ik}\in\mathcal{\bar C}_{k}}\mathsf{g}_{ik}\rho_{k}\|\mathsf{x}_{ik}\|^{-\alpha_k}$ is the out-of-cluster interference.
\end{itemize}

Note that $\interc$ in the denominator of \eqref{eq:sir} is zero when intra-cluster CS is assumed instead of intra-cluster FR. Also, the random variables $\sig$, $\interc$ and $\internc$ are mutually independent.

\section{Coverage Probability}\label{sec:cov_prob}

In this section, the coverage probability is derived for the typical user under non-coherent JT. It is defined as
\begin{IEEEeqnarray}{rCl}
	\pc\triangleq\mathbb{P}\left(\sir\geq\beta\right)
\end{IEEEeqnarray}
for some threshold $\beta>0$. Note that the distributions of $\sig$, $\internc$ and $\interc$ do not exhibit a closed-form expression in general. To get a better handle on the $\sir$ in \eqref{eq:sir}, we therefore propose an approximation of the sum interference $\interc+\internc$ prior to characterizing the $\sir$ for the considered model.

\begin{proposition}[Interference approximation]\label{prop:approx}
The sum interference $\interc+\internc$ in \eqref{eq:sir} can be approximated by a Gamma distributed random variable $\intert$ having distribution $\mathbb{P}(\intert\leq z)=1-\gamma(\nu,z/\theta)/\Gamma(\nu)$, where
\begin{IEEEeqnarray}{rCl}
	\nu=\frac{4\pi\left(\sum_{k}\frac{\lambda_{k}\rho^{-1/\alpha_k}}{\alpha_k-2}\,\mathbb{E}\left[\mathsf{g}_{k}\max\big\{\Delta_k,\tfrac{T_k}{\mathsf{g}_k}\big\}^{1-\frac{2}{\alpha_k}}\right]\right)^2}{\sum_{k}\frac{\lambda_{k}\rho^{-1/\alpha_k}}{\alpha_k-1}\,\mathbb{E}\left[\mathsf{g}_{k}^2\max\big\{\Delta_k,\tfrac{T_k}{\mathsf{g}_k}\big\}^{2-\frac{2}{\alpha_k}}\right]}\label{eq:nu}
\end{IEEEeqnarray}
is the \textit{shape} parameter and
\begin{IEEEeqnarray}{rCl}
	\theta=\frac{\sum_{k}\frac{\lambda_{k}\rho^{-1/\alpha_k}}{\alpha_k-1}\,\mathbb{E}\left[\mathsf{g}_{k}^2\max\big\{\Delta_k,\tfrac{T_k}{\mathsf{g}_k}\big\}^{2-\frac{2}{\alpha_k}}\right]}{2\sum_{k}\frac{\lambda_{k}\rho^{-1/\alpha_k}}{\alpha_k-2}\,\mathbb{E}\left[\mathsf{g}_{k}\max\big\{\Delta_k,\tfrac{T_k}{\mathsf{g}_k}\big\}^{1-\frac{2}{\alpha_k}}\right]}\label{eq:sigma}
\end{IEEEeqnarray}
is the \textit{scale} parameter.
\end{proposition}
\begin{IEEEproof}
Since $\Phi_{k}$ and $\{\mathsf{g}_{ik}\}_{i=0}^{\infty}$ are mutually independent across tiers and by the linearity property of the expectation, the proof follows by computing the mean and variance of $\interc+\internc$ using Campbell's Theorem \cite{stoyan95} and applying a second-order moment-matching, see \cite[Appendix~B]{tanbourgi13_3} for details.
\end{IEEEproof}

For intra-cluster CS in the $k\th$ tier, one has to set $T_k=0$ in  \eqref{eq:nu} and \eqref{eq:sigma}. The Gamma approximation of the sum interference created by Poisson distributed interferers was also previously used in \cite{tanbourgi13_3,ganti09,heath12}, where the accuracy was found satisfactorily high. It can be applied whenever the interference has finite mean and variance.

\begin{theorem}[Coverage probability]\label{thm:cov_prob}
	The coverage probability of the typical receiver in the described HCN setting can be bounded above and below as
	\begin{IEEEeqnarray}{rCl}
		\pc\overset{\tilde\nu=\lfloor\nu\rfloor}{\underset{\tilde\nu=\lceil\nu\rceil}{\lesseqgtr}}1-\sum\limits_{m=0}^{\tilde\nu-1}\frac{(\theta\beta)^{-m}}{m!}\,\frac{\partial^m}{\partial s^{m}}\left[\prod_{k}\mathcal{L}_{\sig_k}(-s)\right]_{s=\frac{-1}{\theta\beta}},\IEEEeqnarraynumspace\label{eq:cov_prob}
	\end{IEEEeqnarray}
where $\mathcal{L}_{\sig_k}(s)$ is given by \eqref{eq:lap_pk} at the top of this page.
\end{theorem}
\setcounter{equation}{8}
\begin{IEEEproof}
See Appendix.
\end{IEEEproof}

\begin{figure*}[!t]
\centerline{\subfloat[]{
	\psfrag{tag1}[c][c]{\small{$\beta$ [dB]}}
	\psfrag{tag2}[c][c]{\small{$\pc$}}
	\psfrag{tag3tag3tag}{\footnotesize{Sim. PPP}}
	\psfrag{tag5}{\footnotesize{Theorem~\ref{thm:cov_prob}}}
	\psfrag{tag6}{\footnotesize{Corollary~\ref{col:approx_pc}}}
	\psfrag{tag7}{\footnotesize{$\max$-$\mathtt{SINR}$}}
	\psfrag{tag8}{\footnotesize{Non-coop}}
	\psfrag{tag9}{\footnotesize{$\alpha=3$}}
	\psfrag{tag10}{\footnotesize{$\alpha=5$}}
	\includegraphics[width=0.49\textwidth]{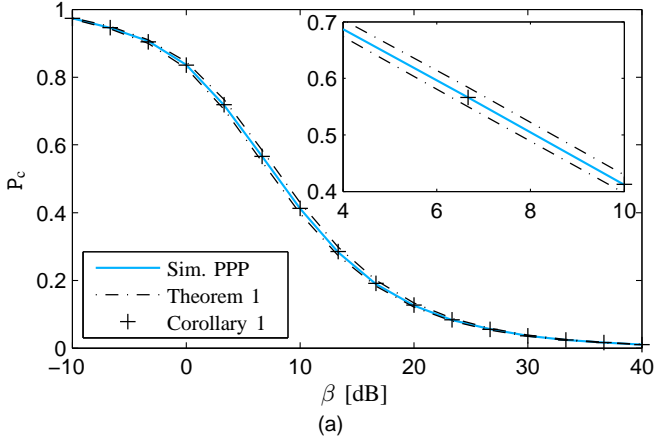}
	\label{fig:sir_valid}}
	\hfil
	\subfloat[]{
	\psfrag{tag1}[c][c]{\small{$\beta$ [dB]}}
	\psfrag{tag2}[c][c]{\small{$\pc$}}
	\psfrag{tag3tag3tag}{\footnotesize{Tier-1 only}}
	\psfrag{tag5}{\footnotesize{Tier-1+Tier-2}}
	\psfrag{tag6}{\footnotesize{All 3 Tiers}}
	\includegraphics[width=0.49\textwidth]{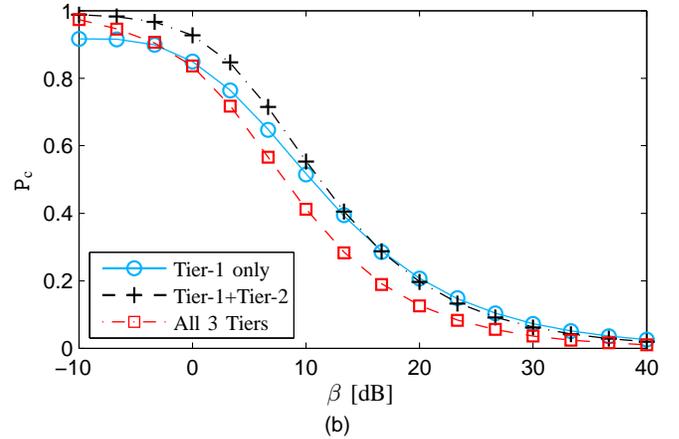}
	\label{fig:sir_tiers}}}
	\caption{Coverage probability $\pc$ vs. $\sir$-threshold $\beta$. Simulation with Poisson interference (solid). Upper/lower bound from Theorem~\ref{thm:cov_prob} (dash-dotted). Linear $\pc$-approximation from Corollary~\ref{col:approx_pc} (``+"-marks). The tier-specific parameters are shown in Table~\ref{tab:parameters}.}\vspace{-.2cm}
\end{figure*}

The worst-case gap between the lower and upper bound is equal to the value of the last summand $m=\unu-1$. For integer-valued $\nu$, either the upper or the lower bound becomes exact. A simple approximation to $\pc$ can be obtained using a linear combination of the bounds in \eqref{eq:cov_prob} with weights chosen according to the relative distance of $\nu$ to $\lnu$ and $\unu$.

\begin{corollary}[Linear approximation of $\pc$]\label{col:approx_pc}
The coverage probability $\pc$ can be approximated as
	\begin{IEEEeqnarray}{rCl}
		\pc&\approx&1-\sum\limits_{m=0}^{\lnu-1}\frac{(\theta\beta)^{-m}}{m!}\,\frac{\partial^m}{\partial s^{m}}\left[\prod_{k}\mathcal{L}_{\sig_k}(-s)\right]_{s=\frac{-1}{\theta\beta}}\IEEEnonumber\IEEEeqnarraynumspace\\
		&&-(\nu-\lnu)\frac{(\theta\beta)^{-\unu+1}}{(\unu-1)!}\,\frac{\partial^{\unu-1}}{\partial s^{\unu-1}}\hspace{-.1cm}\left[\prod_{k}\mathcal{L}_{\sig_k}(-s)\right]_{s=\frac{-1}{\theta\beta}}\hspace{-.4cm}.\IEEEeqnarraynumspace\label{eq:pc_approx}
	\end{IEEEeqnarray}
\end{corollary}

As will be demonstrated later, the approximation in \eqref{eq:pc_approx} turns out to be reasonable accurate despite its simple form. It may furthermore be interesting to study the $\pc$ conditioned upon a fixed number of cooperating BSs in every tier. We denote by $\sig_k|C_k$ the combined received signal power from the $k\th$ tier conditional on $C_k$ cooperative $k\th$-tier BSs.
\begin{corollary}[Conditional Laplace transform of $\sig_k|C_k$]
Conditioned on the fact that $C_{k}$ tier-$k$ BSs belong to the cooperative set of the typical user, the conditional Laplace transform of $\sig_k|C_k$ is
\begin{IEEEeqnarray}{rCl}
	\lap_{\sig_k|C_k}(s)=\left(1+\frac{\Delta_k^{2/\alpha_k}}{\lambda_k\pi\rho_k^{2/\alpha_k}}\log\lap_{\sig_k}(s)\right)^{C_k}.\IEEEeqnarraynumspace\label{eq:lap_pk_con}
\end{IEEEeqnarray}
\end{corollary}

\begin{remark}
Computing the $m$-th derivative in \eqref{eq:cov_prob} is quite involved since \eqref{eq:lap_pk} and \eqref{eq:lap_pk_con} are composite functions. Generally, the $m$-th derivative of composite functions can be efficiently obtained by Fa\`{a} di Bruno's rule and Bell polynomials, given that the derivatives of the outer and inner function are known.
\end{remark}

We next derive the $m$-th derivative of the inner function (i.e., the exponent) of $\lap_{\sig_k}$. The conditional case $\lap_{\sig_k|C_k}$ can be obtained analogously.

\begin{lemma}\label{lem:lap_diff}
	For $m>0$, the $m$-th derivative of the exponent of $\lap_{\sig_k}(-s)$ evaluated at $s=-\frac{1}{\theta\beta}$ is given by
	\begin{IEEEeqnarray}{rCl}
	\frac{\partial^{m}}{\partial s^m}\log\lap_{\sig_k}(-s)\big|_{s=\frac{-1}{\theta\beta}}&=&\frac{2\pi}{\alpha_k}\lambda_k\rho_k^{2/\alpha_k} (\theta\beta)^{m-2/\alpha_k}\IEEEnonumber\\
	&&\hspace{-3cm}\times\mathbb{E}\left[\mathsf{g}_k^{2/\alpha_k}\Gamma\left(m-\tfrac{2}{\alpha_k},\tfrac{\mathsf{g}_k}{\theta\beta}\max\{\Delta_k,\tfrac{T_k}{\mathsf{g}_k}\}\right)\right].\IEEEeqnarraynumspace
	\end{IEEEeqnarray}
\end{lemma}

For the unconditioned case, in particular, the computation of the required derivatives for obtaining \eqref{eq:cov_prob} can be further simplified by exploiting the exponential form of \eqref{eq:lap_pk}.
\begin{corollary}
	The differentiation $\frac{\partial^m}{\partial s^{m}}\left[\prod_{k}\mathcal{L}_{\sig_k}(-s)\right]_{s=-1/\theta\beta}$ for the unconditioned case can be computed by noting that
	\begin{IEEEeqnarray}{rCl}
		\frac{\partial^m}{\partial s^{m}}\left[\prod_{k}\lap_{\sig_k}(-s)\right]_{s=\frac{-1}{\theta\beta}}=
		\frac{\partial^m}{\partial s^{m}}\lap_{\sig}(-s)\Big\lvert_{s=\frac{-1}{\theta\beta}}\IEEEeqnarraynumspace
	\end{IEEEeqnarray}
	where the outer function of $\lap_\sig$ is $e^{x}$ and the inner function has derivative
	\begin{IEEEeqnarray}{rCl}
	\frac{\partial^{m}}{\partial s^m}\log\lap_{\sig}(-s)\big|_{s=\frac{-1}{\theta\beta}}&=&2\pi\sum\limits_{k=1}^{K}\frac{\lambda_k\rho_k^{2/\alpha_k}}{\alpha_k} (\theta\beta)^{m-2/\alpha_k}\IEEEnonumber\\
	&&\hspace{-2cm}\times\mathbb{E}\left[\mathsf{g}_k^{2/\alpha_k}\Gamma\left(m-\tfrac{2}{\alpha_k},\tfrac{\mathsf{g}_k}{\theta\beta}\max\{\Delta_k,\tfrac{T_k}{\mathsf{g}_k}\}\right)\right],\IEEEeqnarraynumspace
	\end{IEEEeqnarray}
where $m>0$.
\end{corollary}

\section{Discussions and Numerical Examples}
We now discuss the results obtained in Section~\ref{sec:cov_prob}, in particular the accuracy of the linear approximation from Corollary~\ref{col:approx_pc}. Numerical examples and design questions are also treated here.

\begin{table}[t]
\footnotesize
\renewcommand{\arraystretch}{1.3}
\caption{HCN Parameters used for Numerical Examples}
\label{tab:parameters}
\centering
\begin{tabular}{c||c|c|c}
\hline
Parameter	&	Tier-1 & Tier-2	&	Tier-3\\
\hline
BS density $\lambda_k$	& $4\,\text{BS/km}^2$ &$16\,\text{BS/km}^2$ &$40\,\text{BS/km}^2$\\
BS power $\rho_k$ & $46\,\text{dBm}$	& $30\,\text{dBm}$ &$24\,\text{dBm}$\\
Path loss $\alpha_k$	&	4.3 & 3.8 & 3.5\\
Nakagami-$m_k$ &	1.8 & 2.3 &2.7\\
Clus.-thres. $\Delta_k$	& -69.6 dBm &  -63.1 dBm & -49.5 dBm\\
Sched.-thres. $T_k$	& $\Delta_1$ &  $\Delta_2+3\,\text{dB}$ & $\Delta_3+3\,\text{dB}$\\
\hline 
\end{tabular}\vspace{-.2cm}
\end{table}

\textbf{Validation and accuracy:} Fig.~\ref{fig:sir_valid} shows the $\pc$ for a HCN with $K=3$ for different $\sir$-thresholds $\beta$. The tier-specific parameters are summarized in Table~\ref{tab:parameters}. The chosen clustering thresholds $\Delta_k$ correspond to an average number of cooperative BSs $\mathbb{E}\left[C_1\right]=3$, $\mathbb{E}\left[C_2\right]=4$ and $\mathbb{E}\left[C_3\right]=2$ in the PPP model. It can be seen that the Gamma approximation of the interference from Proposition~\ref{prop:approx} is accurate as the gap between the lower and upper bound enclosing the simulated $\pc$ is fairly small. Also, the simple approximation from Corollary~\ref{col:approx_pc} performs remarkably well (here, the shape is $\nu=8.5$). 

\textbf{Effect of adding more tiers:} Fig.~\ref{fig:sir_tiers} shows the impact on $\pc$ when adding additional tiers. Interestingly, indicating the performance of non-coherent JT in terms of the number of tiers is not straightforward. For instance, the $\pc$ for Tier-1+Tier-2 HCNs can be higher than for the case of three tiers. This is because, in this example, the clustering threshold $\Delta_3$ in Tier-3 was chosen relatively high, e.g., due to complexity and overhead constraints, resulting in a rather unfavorable ratio of interference and cooperation. Hence, adding more tiers exhibits a non-monotonic trend in terms of $\pc$.   

\begin{figure*}[!t]
\centerline{\subfloat[]{
	\psfrag{tag1}[c][c]{\small{$\Delta_2/\Delta_2'$ [dB]}}
	\psfrag{tag2}[c][c]{\small{$\delta_2$ [\%]}}
	\psfrag{tag3}[c][c]{\small{$\mathbb{E}[\mathtt{R}]$}}
	\psfrag{tag6tag6tag6tag6tag6tag6}{\footnotesize{Av. spectral efficiency $\mathbb{E}[\mathtt{R}]$}}
	\psfrag{tag7}{\footnotesize{Rel. load increase $\delta_2$}}
	\psfrag{tag3tag3tag3tag}{\footnotesize{$T_2/T_2'=-3$ dB}}
	\psfrag{tag4}{\footnotesize{$T_2/T_2'=0$ dB}}
	\psfrag{tag5}{\footnotesize{$T_2/T_2'=3$ dB}}
	\includegraphics[width=0.49\textwidth]{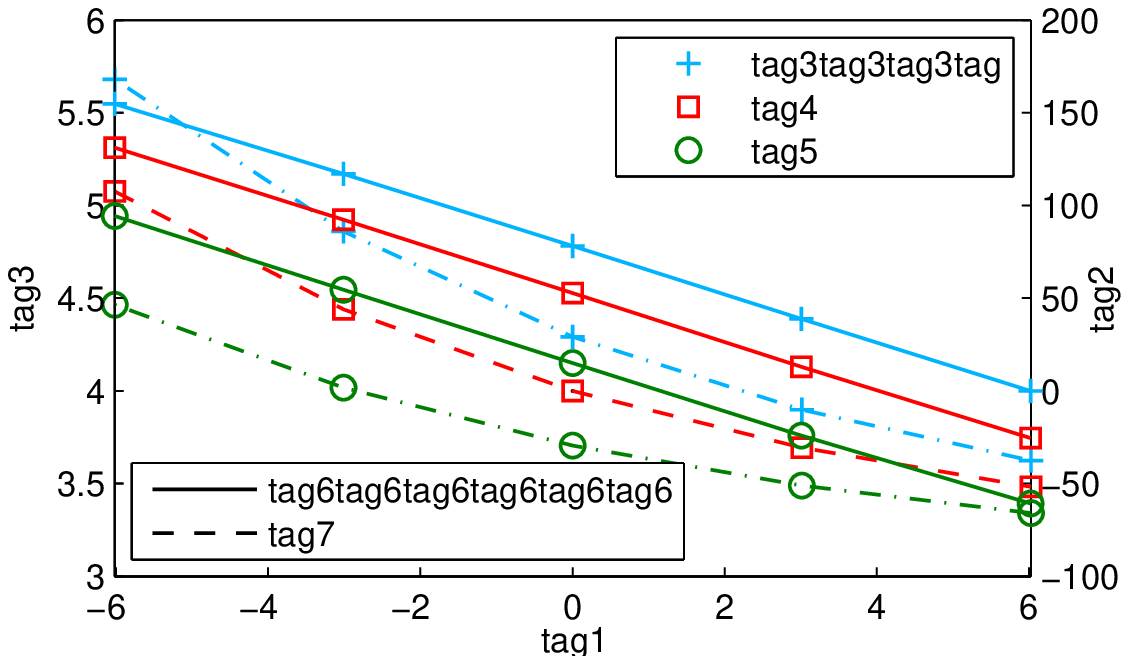}
	\label{fig:sir_exp_se_load}}
	\hfil
	\subfloat[]{
	\psfrag{tag1}[c][c]{\small{\parbox{2cm}{$\tau$ [bit/s/Hz]}}}
	\psfrag{tag2}[c][c]{\small{$\mathbb{P}(\mathtt{R}\leq\tau)$}}
	\psfrag{tag3}{\footnotesize{\parbox{3.2cm}{$\gamma=5.4\%$, $\mathbb{E}[\mathtt{R}]$-loss 0.3\%\\$T_2=\Delta_2-3\,\text{dB}$}}}
	\psfrag{tag4}{\footnotesize{\parbox{3.2cm}{$\gamma=35.2\%$, $\mathbb{E}[\mathbb{R}]$-loss 6\%\\ $T_2=\Delta_2+3\,\text{dB}$}}}
	\psfrag{tag5}{\footnotesize{\parbox{3.4cm}{$\gamma=54.3\%$, $\mathbb{E}[\mathtt{R}]$-loss 14.6\%\\ $T_2=\Delta_2+6\,\text{dB}$}}}
	\includegraphics[width=0.472\textwidth]{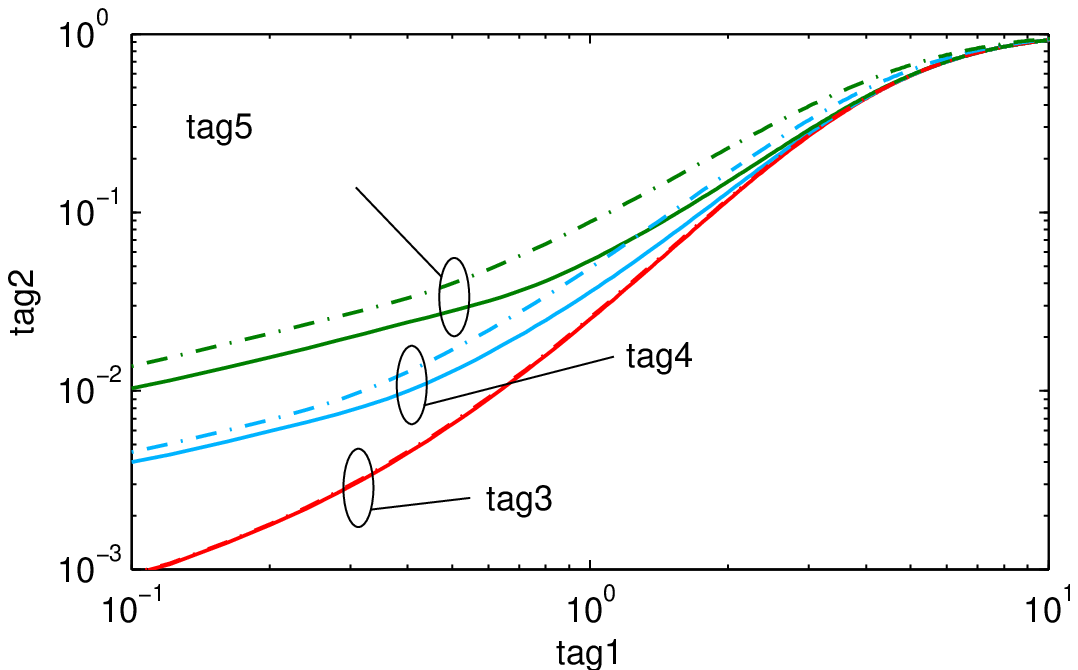}
	\label{fig:se_cs_fr_load}}}
	\caption{Two-tier HCN: (a) Average $\mathtt{R}$ and relative load increase $\delta_2$ for different $\Delta_2$, $T_2$. (b) Distribution of $\mathtt{R}$ for intra-clustering CS (solid line) and FR (dashed line) for different $T_2$. Also shown are the resource saving $\gamma_2$ in the small cells and the $\mathbb{E}[\mathtt{R}]$-loss when switching from CS to FR.}
\end{figure*}

\textbf{Effect of load balancing:} Non-coherent JT can be used also for load balancing, which is especially important in HCNs to avoid under-/over-utilization of the different tiers. Due to transmit power imbalance between the different tiers, this typically means to push users towards smaller cells, e.g., by biasing cell association \cite{singh13}. Balancing load using non-coherent JT is done by varying $\Delta_k$, $T_k$ of the corresponding small cells. Importantly, imprudently stimulating more cooperation by lowering $\Delta_k$ and/or $T_k$ increases the $\sir$, however, possibly at the cost of an overwhelming load increase in the participating small cells. Using the developed model, this effect is analyzed next for the example of a $2$-tier HCN. Since describing cell load in HCNs with cooperation is analytically difficult \cite{tanbourgi13_3}, we use a simple model for characterizing the load increase in the $k\th$-tier cell due to cooperation. Using \eqref{eq:cluster} it can be seen that users closer than $(\Delta_k/\rho_k)^{-1/\alpha_k}$ to a $k\th$-tier BS request cooperation from that BS. Second, given the stationarity of the user point process, the number of radio resources $N$ used for cooperation in a Tier-2 cell is proportional to the number of cooperation requests. Third, fixing an $N'$ through some $\Delta_k'$, $T_k'$, the load increase relative to $N'$ measured as a function of $\Delta_k$, $T_k$ can then be defined as
\begin{IEEEeqnarray}{rCl}
	\delta_k\triangleq \frac{\mathbb{E}\left[N(\Delta_k,T_k)-N(\Delta_k',T_k')\right]}{\mathbb{E}\left[N(\Delta_k',T_k')\right]}.\IEEEeqnarraynumspace\label{eq:load}
\end{IEEEeqnarray}
Applying Campbell's Theorem \cite{stoyan95,HaenggiBook} for evaluating the expectations in \eqref{eq:load}, we obtain
\begin{IEEEeqnarray}{rCl}
 \delta_k=\frac{\mathbb{E}\left[\min\{\Delta_k,T_k/\mathsf{g}_{k}\}^{-2/\alpha_k}\right]}{\mathbb{E}\left[\min\{\Delta_k',T_k'/\mathsf{g}_{k}\}^{-2/\alpha_k}\right]}-1.
\end{IEEEeqnarray}
\begin{remark}
 Note that \eqref{eq:load} does not characterize the total cell load, but rather characterizes the underlying trend as a function of $\Delta_k$ and $T_k$, as these two parameters strongly influence the number of radio resources used for cooperation.   
\end{remark}

Fig.~\ref{fig:sir_exp_se_load} shows how the average spectral efficiency $\mathbb{E}[\mathtt{R}]$ and the relative load increase $\delta_2$ in Tier-2 behave when varying $\Delta_2$, $T_2$. The average spectral efficiency was obtained using the relation $\mathtt{R}\triangleq\log_2(1+\sir)$ and $\mathbb{E}[\mathtt{R}]=\int_{0}^{\infty}\pc(2^{\beta}-1)\,\mathrm d\beta$. It can be seen that $\mathbb{E}[\mathtt{R}]$ increases approximately logarithmically (linearly) as $\Delta$ (in dB) decreases. At high $\Delta_2$, the load increase remains small, but rapidly accelerates as more clustering is stimulated through lowering $\Delta_2$. Interestingly, lowering the activation threshold $T_2$ does not change the load significantly at higher $\Delta_2$ while providing considerable spectral efficiency gains. An important insight is that balancing load, by stimulating small cells (decreasing $\Delta_2$, $T_2$) to assist macro BSs through non-coherent JT, may be favorable only to a certain extent since the additional load imposed on small cells eventually outpaces the spectral efficiency gains.

\textbf{Intra-cluster scheduling:}
An important design question in non-coherent JT is whether cooperating BSs not participating in an ongoing cooperative transmission 1) should reuse the radio resources allocated to non-coherent JT (intra-cluster FR) or 2) should remain silent on these resources to avoid intra-cluster interference, thereby virtually increasing cell load in these cells (intra-cluster CS). By trading off intra-cluster interference and cell load against cooperation, the impact of the two scheduling schemes is moreover intensified by the activation threshold $T_k$. In HCNs, in particular, intra-cluster FR might be favorable in small cells to obtain ``cell-splitting'' gains. 

We next study whether intra-cluster FR or intra-cluster CS should be used in smaller cells using the following metric: switching from CS to FR invokes a resource saving at cooperative $k\th$-tier BSs not participating in non-coherent JT. This saving directly translates into a load reduction at those BSs, which we characterize as
\begin{IEEEeqnarray}{rCl}	\gamma_k\triangleq1-\mathbb{E}\left[\frac{\sum_{\mathsf{x}_{ik}\in\mathcal{C}_k}\mathds{1}\left(\mathsf{g}_{ik}\rho_k\|\mathsf{x}_{ik}\|^{-\alpha_k}\geq T_k\right)}{\sum_{\mathsf{x}_{ik}\in\mathcal{C}_k}\mathds{1}(\mathsf{x}_{ik}\in\mathcal{C}_k)}\right],\IEEEeqnarraynumspace\label{eq:gamma}
\end{IEEEeqnarray}
i.e., the \emph{spatially}-averaged radio resource saving in cooperative $k\th$-tier cells of the typical user. The load reduction in \eqref{eq:gamma} can be computed as
\begin{IEEEeqnarray}{rCl}
	\gamma_k=1-\mathbb{E}\left[\min\left\{1,(\mathsf{g}_k\Delta_k/T_k)^{2/\alpha_k}\right\}\right].
\end{IEEEeqnarray}
Interestingly, $\gamma_k$ does not depend on $\lambda_k$ and $\rho_k$. Fig.~\ref{fig:se_cs_fr_load} shows the distribution of $\mathtt{R}$ for the example of a two-tier HCN for different $T_2$. The value of $\Delta_2$ was chosen such that $\mathbb{E}[C_2]=5$. It can be seen that at low $T_2$, switching from CS to FR does barely affect $\mathtt{R}$ (or $\mathbb{E}[\mathtt{R}]$) while a load saving of approximately $5.4$\% is achieved. In this regime, FR may thus be more favorable. For larger $T_2$ one has to bite the bullet: much higher savings, e.g., $54.3$\%, can be obtained, however, at the cost of worsening $\mathtt{R}$, e.g., $\mathbb{E}[\mathtt{R}]$-loss  of 14.6\%. In lightly-loaded cells CS should hence be used when a high $T_2$ is desired in order to additionally profit from muting intra-cluster interference.

\section{Conclusion}
We developed a tractable model and derived the coverage probability for non-coherent JT in HCNs, thereby accounting for the heterogeneity of various system parameters including BS clustering, channel-dependent cooperation activation, and radio propagation model. To the best of the authors' knowledge this is the first work to analyze cooperation in such generic HCNs. The developed theory allowed us to treat important design questions related to load balancing and intra-cluster scheduling.


\appendix
\subsection{Proof of Theorem~\ref{thm:cov_prob}}
We write
\begin{IEEEeqnarray}{rCl}
	\pc&=&\mathbb{E}_{\sig}\left[\mathbb{P}\left(\interc+\internc<\sig/\beta\right)\right]\IEEEnonumber\\
		&\overset{\text{Prop.~\ref{prop:approx}}}{\simeq}&1-\mathbb{E}_{\sig}\left[\mathbb{P}\left(\intert\geq\sig/\beta\right)\right]\IEEEnonumber\\
	&=&1-\mathbb{E}_\sig\left[\frac{\Gamma\left(\nu,\sig/\theta\beta\right)}{\Gamma(\nu)}\right].
\end{IEEEeqnarray}
Noting that $\Gamma(a,z)/\Gamma(a)$ is monotone increasing in $a$ for all $z\geq0$, we obtain the inequality
\begin{IEEEeqnarray}{rCl}
	\pc&\overset{\tilde\nu=\lfloor\nu\rfloor}{\underset{\tilde\nu=\lceil\nu\rceil}{\lesseqgtr}}&1-\mathbb{E}_\sig\left[\frac{\Gamma\left(\tilde\nu,\sig/\theta\beta\right)}{\Gamma(\tilde\nu)}\right]\IEEEnonumber\\
	&=&1-\sum_{m=0}^{\tilde \nu-1}\frac{(\theta\beta)^{-m}}{m!}\mathbb{E}_{\sig}\left[\sig^{m}e^{-\sig/\theta\beta} \right]\IEEEnonumber\\
	&=&1-\sum_{m=0}^{\tilde \nu-1}\frac{(\theta\beta)^{-m}}{m!}\,\frac{\partial\mathcal{L}_\sig(-s)}{\partial s^{m}}\Big\lvert_{s=\frac{-1}{\theta\beta}},
\end{IEEEeqnarray}
where $\mathcal{L}_\sig(s)$ is the Laplace transform of the combined received signal power. Due to the independence property of the $\Phi_{1},\ldots,\Phi_{K}$, we can decompose $\mathcal{L}_\sig(s)$ into $\prod_k\mathcal{L}_{\sig_k}(s)$, where $\mathcal{L}_{\sig_k}(s)$ is the Laplace transform corresponding to the received power $\mathsf{P}_{k}$ from tier $k$ BSs. It can be obtained as
\begin{IEEEeqnarray}{rCl}
	&&\mathcal{L}_{\sig_k}(s)=\mathbb{E}\left[\exp\left\{-s\rho_k\sum_{\mathsf{x}_{ik}\in\mathcal{C}_{\text{a},k}}  \mathsf{g}_{ik}\|\mathsf{x}_{ik}\|^{-\alpha_k}\right\}\right]\IEEEnonumber\\
	&&\quad\overset{(a)}{=}\mathbb{E}_{\Phi_k}\Bigg[\prod\limits_{\mathsf{x}_{ik}\in\mathbb{R}^2} \mathbb{E}_{\mathsf{g}_{ik}}\Big[\exp\big\{-s\rho_k\mathsf{g}_{ik}\|\mathsf{x}_{ik}\|^{-\alpha_k}\IEEEnonumber\\
	&&\quad\quad\times\mathds{1}(\mathsf{g}_{ik}\rho_{k}\|\mathsf{x}_{ik}\|^{-\alpha_k}\geq T_k)\,\mathds{1}(\rho_{k}\|\mathsf{x}_{ik}\|^{-\alpha_k}\geq \Delta_k)\big\}\Big]\Bigg]\IEEEnonumber\\
	&&\quad\overset{(b)}{=}\exp\bigg\{-\lambda_{k}\int_{\mathbb{R}^2}1-\mathbb{E}_{\mathsf{g}_{k}}\bigg[\exp\Big\{-s\rho_k\mathsf{g}_{k}\|x\|^{-\alpha_k}\IEEEnonumber\\
	&&\qquad\qquad\qquad\times\mathds{1}\big(\rho_k\|x\|^{-\alpha_k}\geq\max\{\Delta_k,\tfrac{T_k}{\mathsf{g}_{k}}\}\big)\Big\}\bigg]\mathrm dx\bigg\}\IEEEeqnarraynumspace\IEEEnonumber\\
	&&\quad\overset{(c)}{=}\exp\bigg\{-\tfrac{2\pi}{\alpha_k}\lambda_{k}\rho_k^{2/\alpha_k}\IEEEnonumber\\
	&&\qquad\quad\times\mathbb{E}_{\mathsf{g}_{k}}\bigg[\int_{\max\{\Delta_k,\frac{T_k}{\mathsf{g}_{k}}\}}^{\infty}\hspace{-.3cm}t^{-1-2/\alpha_k}\,(1-e^{-s\mathsf{g}_kt})\,\mathrm dt\bigg]\bigg\},\IEEEeqnarraynumspace
\end{IEEEeqnarray}
where (a) follows from the i.i.d. property of the $\mathsf{g}_{ik}$, (b) follows from the probability generating functional of a PPP \cite{stoyan95,HaenggiBook}, and (c) follows from interchanging expectation and integration and from the substitution $t=\rho_k\|x\|^{-\alpha_{k}}$. Eq.~\eqref{eq:lap_pk} then follows after partial integration. \qed

\bibliographystyle{IEEEtran}

\end{document}